\documentclass[aip,graphicx,jcp]{revtex4-1}

\usepackage{upgreek}
\usepackage{mathtools}
\usepackage{xcolor}
\usepackage{soul}

\draft 

\begin{document}


\title{Kinetic rates calculation via non-equilibrium dynamics} 

\author{Bruno Stegani}%
\affiliation{ 
Department of Biosciences, Università degli Studi di Milano, Via Celoria 26, 20133 Milan, Italy.
}

\author{Riccardo Capelli}%
 \email{riccardo.capelli@unimi.it}
\affiliation{ 
Department of Biosciences, Università degli Studi di Milano, Via Celoria 26, 20133 Milan, Italy.
}

\date{\today}

\begin{abstract}
This study introduces a novel computational approach based on ratchet-and-pawl molecular dynamics (rMD) for accurately estimating ligand dissociation kinetics in protein-ligand complexes. By integrating Kramers' theory with Bell's equation, our method systematically investigates the relationship between the effective biasing force applied during simulations and the ligand residence times. The proposed technique is demonstrated through extensive simulations of the benzamidine-trypsin complex, employing first an implicit solvent model (multi-eGO) to set up the approach parameters and thus an explicit solvent model. Our results illustrate the method's reliability, accuracy, and computational efficiency, with calculated kinetic rates closely matching experimental values. Overall, this study highlights rMD as a versatile and efficient non-equilibrium methodology, broadly applicable to kinetic analyses in chemical and biological systems.
\end{abstract}


\maketitle 
\section{\label{intro} Introduction}
A central challenge in molecular simulation is the calculation of protein–ligand unbinding rates, a problem of particular relevance in pharmacology where residence time is closely linked to drug efficacy.\cite{Knockenhauer2023} These processes are often governed by complex, multistep mechanisms, making their quantitative description a demanding task for theory and computation.

The improvement of the available computational capabilities made the \textit{in silico} approach to this class of problems viable, despite the huge gap in molecular dynamics (MD) between the accessible timescales (in the order of the $\mu$s) and the experimental ones (between ms and h). To bridge this gap, numerous different approaches have been developed, aiming at identifying the kinetic rates of ligand binding and unbinding\cite{Bernetti2019,Ahmad2022}, which can be classified into three main classes: (i) biased methods\cite{Tiwary2013,Miao2018,Kokh2018}, (ii) Markov state modeling\cite{Singhal2004,Wu2016}, and (iii) path sampling approaches\cite{Huber1996,Elber2007}. \\
Among those, non-equilibrium biasing techniques are particularly promising for binding kinetics calculations: unlike most other methods, they can capture rare events in only a fraction of the simulation time required by unbiased approaches. Notable examples are $\tau$-RAMD\cite{Kokh2018}, where a randomly directed fixed force drags the ligand out from its binding pocket (for this technique, absolute $k_{\text{off}}$ estimates have been reported\cite{Maximova2021}), elABMD\cite{Gobbo2019} where the kinetics is estimated via non-equilibrium simulations which bias an apt reaction coordinate based on the electrostatic interactions between the ligand and the pocket involved in binding, and PCV-ABMD which used non-equilibrium simulations on different mutants of the receptor with the same ligand to compute relative residence times\cite{Coricello2025}.

Here, we present a kinetic rate calculation approach based on Kramers' theory\cite{Kramers1940} which employs ratchet-and-pawl MD\cite{Marchi1999,Camilloni2011} (rMD). We show that there is a linear dependence between the effective force $F_{\text{eff}}$ applied by rMD and the double logarithm of the residence time of the ligand $\tau$, similar to what has been shown for another non-equilibrium technique\cite{Maximova2021}. Performing a series of independent (and thus highly parallelizable) simulations with different harmonic bias constants $k^{\text{rMD}}$, we were able to compute the unbinding rate of the benzamidine-trypsin complex represented with an \textit{in vacuo} structure-based model (thus extremely fast) and an atomistic, explicit-solvent model. We took advantage of the possibility of obtaining the structure-based force field unbinding $k_{\text{off}}$ to calibrate some parameters of the approach to exploit them on the explicit-solvent one. As a result, we are in good agreement with the known value $k_{\text{off}}$. The computational cost is comparable to the state-of-the-art techniques used on the same system (e.g., On-the-fly Probability Enhanced Sampling\cite{Ansari2022,Ray2022}, Adaptive Multilevel Splitting\cite{Teo2016}).

\section{\label{methods}Methods}
\subsection{Theoretical framework}
In rMD, the biasing potential as a function of a collective variable (CV) $s(t)$ is defined as 
\begin{equation}
    V_{\text{rMD}}(s(t))=
    \begin{cases}
        \displaystyle\frac{k^{\text{rMD}}}{2} \left( s(t) - s_{\max}(t) \right)^2 ~~~~\text{if } s(t)<s_{\text{max}}(t) \\
        0 ~~~~~~~~~~~~~~~~~~~~~~~~~~~~~~~~\text{if } s(t)\geq s_{\text{max}}(t)
    \end{cases},
\end{equation}
where $k^{\text{rMD}}$ is the harmonic constant and $s_{\text{max}}(t)$ is the maximum value of the CV observed in the simulation. The potential acts as a ratchet-and-pawl mechanism, discouraging the return of the system at the initial CV values. One of the key points of this technique is given by the fact that it is adiabatic in the CV space, i.e. the work sums at zero when the system overcomes $s_{\text{max}}$. \\
Although we do not have a constant force performing work along the CV, we can similarly interpret the effective force $F_{\text{eff}}$, defined as the average force applied by rMD throughout the trajectory. Kramers' theory\cite{Kramers1940} can thus be used to estimate the kinetics of the transition driven by rMD.
The effect of a dragging force $F_\text{eff}$ can be taken into account using Bell's equation\cite{Bell1978},
\begin{equation}
\label{eq:bell}
    \tau = \nu_{0} \exp\left(\frac{E_{0} - \gamma F_{\text{eff}} }{k_{B} T}\right),
\end{equation}
where $\nu_{0}$ represents the fluctuations in the bound state (the reciprocal of the natural frequency), $E_{0}$ is the binding energy, $\gamma$ is the effective distance over which the force does work, $k_B$ and $T$ are Boltzmann’s  constant and system’s temperature, respectively. \\
In a rMD run with a reasonably small $k^{\text{rMD}}$, the system progresses quasi-adiabatically (\textit{i.e.}, without the injection of a large amount of energy from the biasing potential) through a sequence of metastable states defined by the moving barrier. In this regime, each local free-energy barrier is crossed predominantly by thermal fluctuations, with a probability $\propto e^{-E/k_B T}$. If the effective distance $\gamma$ reflects the width of such a barrier, and the energetic cost of widening it grows linearly with $\gamma$, then the Boltzmann factor implies that the occurrence probability of a given $\gamma$ decays exponentially. This directly yields
\begin{equation}
\label{eq:gamma_rmd}
    \gamma = \gamma_{0} \exp \left( -\frac{F_{\text{eff}}}{F_{c}} \right),
\end{equation}
where $F_c$ is the characteristic force scale quantifying how rapidly the biasing force reduces the effective width of the dissociation barrier.

We can now define a rate equation
\begin{equation}
\label{eq:rate_rmd}
    \frac{d\tau}{d F_{\text{eff}}} = -\frac{\gamma}{k_{B} T} \tau,
\end{equation}
and, inserting (\ref{eq:gamma_rmd}) in (\ref{eq:rate_rmd}) and solving we have
\begin{equation}
\label{eq:tau_rmd}
    \tau = \tau_0 \exp \left( -\frac{\gamma_0 F_c}{k_{B}T} \left( 1- \exp \left( -\frac{F_{\text{eff}}}{F_{c}} \right) \right) \right)
\end{equation}
which, for $F_{\text{eff}}=0$, returns as expected $\tau=\tau_{0}$. \\
Taking the logarithm of both sides, and then the logarithm again,we can obtain a solution which explicitly involves $\log(\log \tau_0)$:
\begin{widetext}
\begin{equation}\label{eq:doublelog}
    \log(\log (\tau)) = \log(\log (\tau_0)) + \log\left( 1 - \frac{\gamma_0 F_c}{k_B T \log (\tau_0)} \left( 1 - \exp\left(-\frac{F_{\text{eff}}}{F_c} \right) \right) \right).
\end{equation}
\end{widetext}
For small forces, such that $\exp(-F_{\text{eff}} / F_c) \ll 1$, the logarithmic term can be linearized using $\log(1 - \varepsilon) \approx -\varepsilon$ for small $\varepsilon$. This leads to the approximate linear form
\begin{equation}
    \log(\log (\tau)) \approx \log(\log (\tau_0)) - \frac{\gamma_0}{k_B T \log (\tau_0)} F_{\text{eff}}.
\end{equation}
We stress here that Eq.~(7) is obtained under the assumption of weak effective forces, \textit{i.e.,} $F_{\text{eff}} \ll F_c$. In this regime the logarithmic term can be safely linearized, yielding the observed doubly logarithmic dependence of $\tau$. For larger forces, the full expression of Eq.(\ref{eq:tau_rmd}) should be retained. \\
This doubly logarithmic dependence allows for a straightforward linear extrapolation of $\tau$ to the zero-force limit, where $\log(\log (\tau))$ is expected to converge to $\log(\log (\tau_0))$. This is in analogy with a previous work by Maximova \textit{ et al.}, where they found a similar dependence for non-equilibrium dynamics driven by $\tau$-RAMD\cite{Kokh2018}.

The double exponential form in equation \eqref{eq:tau_rmd} is not accidental. We can write the survival probability $S$ (except for a normalization factor), defined as the probability that the ligand has not yet dissociated after time $\tau$, as 
\begin{equation}
    S(\tau) = \exp \left( -\frac{\tau}{\tau_0} \exp \left(\kappa(1- e^{-F_{\text{eff}}/F_c}) \right)  \right)
\end{equation}
where $\kappa = \frac{\gamma_0 F_c}{k_B T}$ is a constant. Intuitively, this links to the Gumbel distribution\cite{Gumbel1941}, which describes the distribution of minima or maxima in random variables. Thus, the dissociation event in rMD simulations behaves statistically like an extreme fluctuation problem. In general, the overcoming of the free energy barrier is driven by thermal fluctuations, but dissociation is primarily determined by the largest fluctuation occurring along the trajectory (an effect also observed in disordered systems\cite{Bouchaud1997}). Such extreme fluctuations are selected by the ratchet-and-pawl mechanism of rMD, which favors the selection of extreme events in the direction of the transition, which explains why the double exponential form appears.

\subsection{Multi-eGO Simulations}
Multi-eGO\cite{Scalone2022,BacicToplek2023,Stegani2024} is a quasi-atomistic (it includes only heavy atoms) structure-based model with implicit solvent and transferable bonded interactions. To define the non-bonded interactions for a protein in this framework we need to train the system on a simulation performed with a realistic potential (in this case DES-Amber\cite{Piana2020}, an all-atom explicit solvent classical force field) and a prior potential, called random coil, which contains only hardcore repulsion and bonded terms that accounts for the ``trivial'' interaction (i.e., the ones that appear for geometric vicinity due to chemical bonding). By combining the probability distributions using a bayesian approach, we can obtain the interaction energy for every atom pair $\varepsilon_{ij}$ as
\begin{equation}
\label{eq:mego_eps}
    \varepsilon_{ij} = - \frac{\varepsilon_{0}}{\log(P^{\text{RC}}_{\text{thr}})} \log \left( \frac{P^{\text{MD}}_{ij}}{\max\left( P^{\text{RC}}_{ij},P^{\text{RC}}_{\text{thr}}  \right) } \right),
\end{equation}
where $P^{\text{MD}}_{ij}$ is the probability distribution of the distance between the atom $i$ and $j$ in the atomistic simulation, $P^{\text{RC}}_{ij}$ is the probability distribution of the distance between the atom $i$ and $j$ in the random coil simulation, $P^{\text{RC}}_{\text{thr}}$ is the minimum probability used for regularization, and $\varepsilon_{0}$ is the energy scale of the interaction. 
To set the value of $\varepsilon_0$, we compare the Root Mean Square Fluctuations (RMSF) and the Gyration radius distribution ($R_G$) of the all-atom and multi-eGO simulations with respect to the energy scale. \\ 
To obtain a multi-eGO potential for trypsin, we performed the following training simulations:
(i) Five replicates of 1-$\mu$s-long all-atom simulations in explicit solvent of trypsin alone (same parameters and force field of the all-atom simulations detailed in the next section), and (ii) a 1 $\mu$s-long simulation of trypsin alone with a random coil potential. \\
Analyzing the resulting potential at different energy scales, we set $\varepsilon_{0}= 0.35$ kJ/mol (see details in the Supporting Information). 

For protein-ligand interaction, we follow a similar approach: we perform (i) a detailed simulation with the protein and a set of 5 ligands (one in its binding pocket, four in the solvent), and (ii) a random coil simulation, which in this case is represented by the multi-eGO representation of the protein with the same number of solvated ligands of the detailed reference simulation immersed in a simulation box of the same size. In the latter simulation, we do not have any attractive interaction between the protein and the ligand; in this way, we can have information regarding random collisions between the ligand and the surface of the protein, which can be inserted in the same bayesian procedure shown in Eq. (\ref{eq:mego_eps}), thus removing rototranslational entropy and returning a meaningful estimate of the protein-ligand interaction. Similarly to what we did for the intramolecular interaction in trypsin, we have to find the intermolecular $\varepsilon^{\text{inter}}_{0}$ for the protein-ligand interactions by comparing the contact probability between the ligand and the protein with respect to the reference simulation varying the value of $\varepsilon^{\text{inter}}_{0}$. After considering multiple values of the energy scale, we set $\varepsilon^{\text{inter}}_{0}=0.41$ kJ/mol (see details in the Supporting Information). 

In the unbiased MD simulations to compute the $k^{\text{MD}}_{\text{off}}$ we performed 10 different replicates of a 10 $\mu$s-long simulation with different starting velocities. After completion of the simulations, we calculated the observed residence times $t$, fitting the empirical cumulative distribution to a Poisson distribution $P(t) = 1-\exp(-t/\tau)$, and finally comparing them with a Kolmogorov-Smirnov test\cite{Salvalaglio2014} (see Figure S1).

For the rMD simulations, we performed 660 simulations initializing the system with the benzamidine in its bound state with different velocity distributions. For every $k^{\text{rMD}}$ values, ranging from 1 to 100 kJ/mol/nm$^2$, we performed 20 independent simulations (complete list in Table S2).

The simulation parameters were the same for random coil and multi-eGO simulations (unbiased and rMD): employed a Langevin dynamics integrator in the NVT ensemble with a timestep of 5 fs, van der Waals cutoff of 1.44 nm, neighborlist cutoff of 1.58 nm, and a thermal time coupling of 25 ps to keep the temperature at 300 K. All the simulations have been performed with GROMACS 2024.4\cite{Abraham2015} patched with PLUMED 2.9.3\cite{Tribello2014,plumed2019}. 

\subsection{All-atom Simulations}
To prepare the atomistic model of the benzamidine-trypsin complex, we started from the crystallographic structure 3PTB\cite{Marquart1983} deposited in the Protein Data Bank. We parameterized the protein using the DES-Amber force field\cite{Piana2020}, the ligand using GAFF2\cite{Wang2004} and computing single-point charges with AM1-BCC semiempirical model\cite{Jakalian2000,Jakalian2002}. We inserted the protein-ligand model in a cubic box (8.5 nm side) and solvated in the TIP4P-D water model\cite{Piana2015}. The protonation states for titrable residues have been determined by Schr\"odinger Maestro 2023.3 Protein Preparation Wizard for a pH value of 7.4, and the box has been neutralized by adding 8 Cl\textsuperscript{-} ions (one chlorine ion was already present in the experimental structure). The choice of a zero ionic force with respect to a more physiological concentration of NaCl should not affect the value of $k_{\text{off}}$ for the benzamidine-trypsin complex, as shown by Votapka and coworkers\cite{Votapka2017}.

We initially minimized the structure with 1000 steps of steepest descent minimization, followed by 1000 steps of conjugate gradient minimization. We then performed 2 ns of NPT simulations with position restraints ($k = 1000$ kJ/mol/nm$^2$ in all directions) on all the heavy atoms of trypsin and benzamidine, followed by 500 ps of relaxation removing the position restraints.

Similarly to the multi-eGO simulations, we performed 600 rMD simulations initializing the system with the benzamidine in its bound state with different velocity distributions. For every $k^{\text{rMD}}$ values, ranging from 16 to 1000 kJ/mol/nm$^2$, we performed 20 independent simulations (complete list in Table S3).

All simulations were performed with a 2 fs time step. The cutoff for van der Waals and short-range interactions was set to 1 nm, and long-range electrostatic interactions were calculated using the Ewald smooth particle mesh method\cite{Essmann1995}, Temperature control was achieved using the velocity rescale thermostat\cite{Bussi2007} at
300 K (coupling separately the protein, the ligand, and the water and ion group, with a coupling time of 1 ps), and pressure was maintained at 1 bar using the isotropic cell rescale barostat\cite{Bernetti2020} (water compressibility set at $4.5\cdot10^{-5}$ bar$^{-1}$, and a coupling time of 5 ps).
All simulations were performed with GROMACS version 2024.4\cite{Abraham2015} patched with PLUMED 2.9.3\cite{Tribello2014,plumed2019}.

\subsection{Data filtering and Error Estimation}
To compute the time-average of the applied force and determine $F_{\text{eff}}$, we first needed to establish the point in the CV space at which the ligand is considered unbound. This choice is crucial because, beyond this point, subsequent segments of the trajectory experience only the entropic force from the solvent ---or, in the case of the multi-eGO model, lack solvent altogether--- which would otherwise lower the $F_{\text{eff}}$ value. Obviously, such choice can be performed \textit{a posteriori}, being sure that in the simulation the nonequilibrium process is completed (i.e., imposing a large value of the CV to stop the simulation, in our case 3.2 nm). After analyzing several unbinding trajectories, we found that a CV value of 1.1 nm indicates that the ligand is completely solvated. Consequently, we computed the average using only the portions of the trajectories that fall below this threshold.

To distinguish between adiabatic and inertial trajectories, we employed the Density Peak Clustering (DPC) algorithm\cite{Rodriguez2014}. The input consisted of the time series of the CV for each replica, based on the intuition that the temporal evolution of the reaction coordinate differs markedly between the two regimes. To ensure a common metric, we standardized the length of each time series to 100,000 points. For shorter series, we applied Savitzky-Golay filtering\cite{Savitzky1964} (window length=11, polynomial order=3) to smooth and fill in data, whereas for longer series, we employed uniform subsampling. In the decision graph, we constrained the algorithm to identify two clusters, thereby differentiating the two classes of trajectories.

A second filtering step—again using DPC—was then applied to remove points exhibiting inertial behavior at low $k^{\text{rMD}}$ values. These outliers are not detectable in the first filtering because, at large $k^{\text{rMD}}$, the behavior of inertial trajectories becomes very similar to that of adiabatic ones. In this step, the clustering algorithm was executed on coordinates plotted in the $\log(\log(\tau))/F_{\text{eff}}$ plane. To avoid bias due to the differing orders of magnitude, we rescaled both dimensions to the range (0, 1). To eliminate points that lie far from regions of maximum density (i.e., outliers), we set the number of clusters to one and defined a density threshold, $\rho_{\text{thr}}$, to separate inlier points from noise. 
To define such threshold we consider the multi-eGO simulations, comparing the $k^{\text{MD}}_{\text{off}}$ obtained via long MD simulations to $k^{\text{rMD}}_{\text{off}}$ computed varying the value of the $\rho_{\text{thr}}$ and performing a linear fit to get the residence time at zero-force $\tau_{0}$. the values of $k^{\text{MD}}_{\text{off}}$ and $k^{\text{rMD}}_{\text{off}}$ matched at $\rho_{\text{thr}}= 0.3$, and we kept this value also for the all-atom filtering. We determined this threshold by comparing $k^{\text{MD}}_{\text{off}}$, obtained from long MD simulations, to $k^{\text{rMD}}_{\text{off}}$ calculated at various $\rho_{\text{thr}}$ values while performing a linear fit to extrapolate the zero-force residence time, $\tau_{0}$. The match between $k^{\text{MD}}_{\text{off}}$ and $k^{\text{rMD}}_{\text{off}}$ occurred at 
$\rho_{\text{thr}}=0.3$, so we adopted this value for all-atom filtering.

For all the linear fits performed, we computed the confidence interval by carrying out 10,000 bootstrap instances to compute the 95\textsuperscript{th} percentile of the calculated values.

\section{\label{results}Results and Discussion}
\subsection{Multi-eGO Model}
The first model of the benzamidine-trypsin complex on which we applied our protocol was built using a framework called multi-eGO\cite{Scalone2022,BacicToplek2023,Stegani2024}. We initially built the potential using explicit-solvent atomistic simulations as a training set (see Methods and the Supporting Information). Multi-eGO potentials preserve the equilibrium properties of protein-ligand interactions while speeding up the kinetics of events, which is mainly caused by the removal of the solvent contributions to the system entropy\cite{Stegani2024}.Exploiting this acceleration of the system time scales, we initially performed long unbiased MD simulations (10 replicas, 10 $\mu$s-long each starting from the benzamidine-trypsin complex in its bound state) to evaluate the $k_{\text{off}}$ of the multi-eGO model. Analyzing these simulations we observed 40 spontaneous unbinding events that we fitted to a Poisson distribution to get the residence time value ($\tau$), verifying, obtaining a p-value $>$ 0.15 via a Kolmogorov-Smirnov test that we observed a rare event\cite{Salvalaglio2014} (p-value 0.636, Figure S1). The resulting $\tau$ is 2.4 $\mu$s ([1.5, 3.6] 95\% CI), which corresponds to a $k^{\text{MD}}_{\text{off}}=4.2\cdot 10^{5}$ s$^{-1}$ ($[2.8\cdot 10^{5},6.7\cdot 10^{5}]$ s$^{-1}$ 95\% CI), about three orders of magnitude faster than the experimental value\cite{Guillain1970} ($k^{\text{exp}}_{\text{off}}=(6 \pm 3) \cdot 10^{2}\,\text{s}^{-1}$).

\begin{figure}[h!]
    \includegraphics[width=0.45\textwidth]{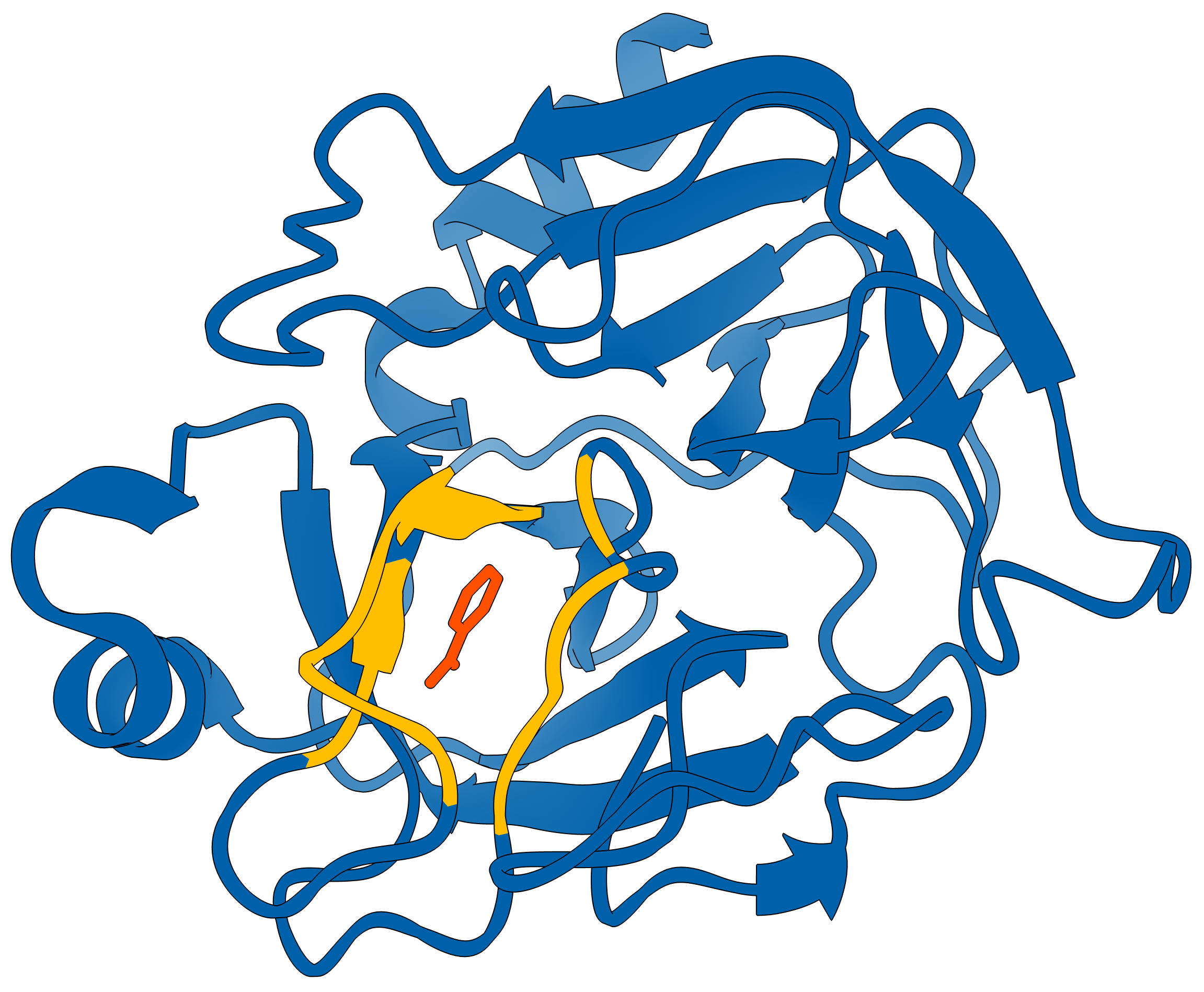}
    \caption{Cartoon representation of benzamidine-trypsin complex. In blue, the residues of trypsin farther than 5 \AA{} from benzamidine, in yellow the amino acids that we selected to define the distance between the pocket and the ligand, and in orange the benzamidine molecule in its binding site.}
    \label{img:render}
\end{figure}

Having computed a ground truth value for the system through unbiased molecular dynamics, we set up our rMD simulation campaign. First, we have to define an appropriate collective variable (CV) that can be used to push the system from the bound to the unbound state. We considered all trypsin residues closer than 5 \AA{} from any atom of benzamidine (see Table S1), and we defined as CV the distance between the C$_{\alpha}$ atom of such amino acids and the center of mass of benzamidine (see Figure \ref{img:render}).

\begin{figure}[h!]
    \includegraphics[width=0.45\textwidth]{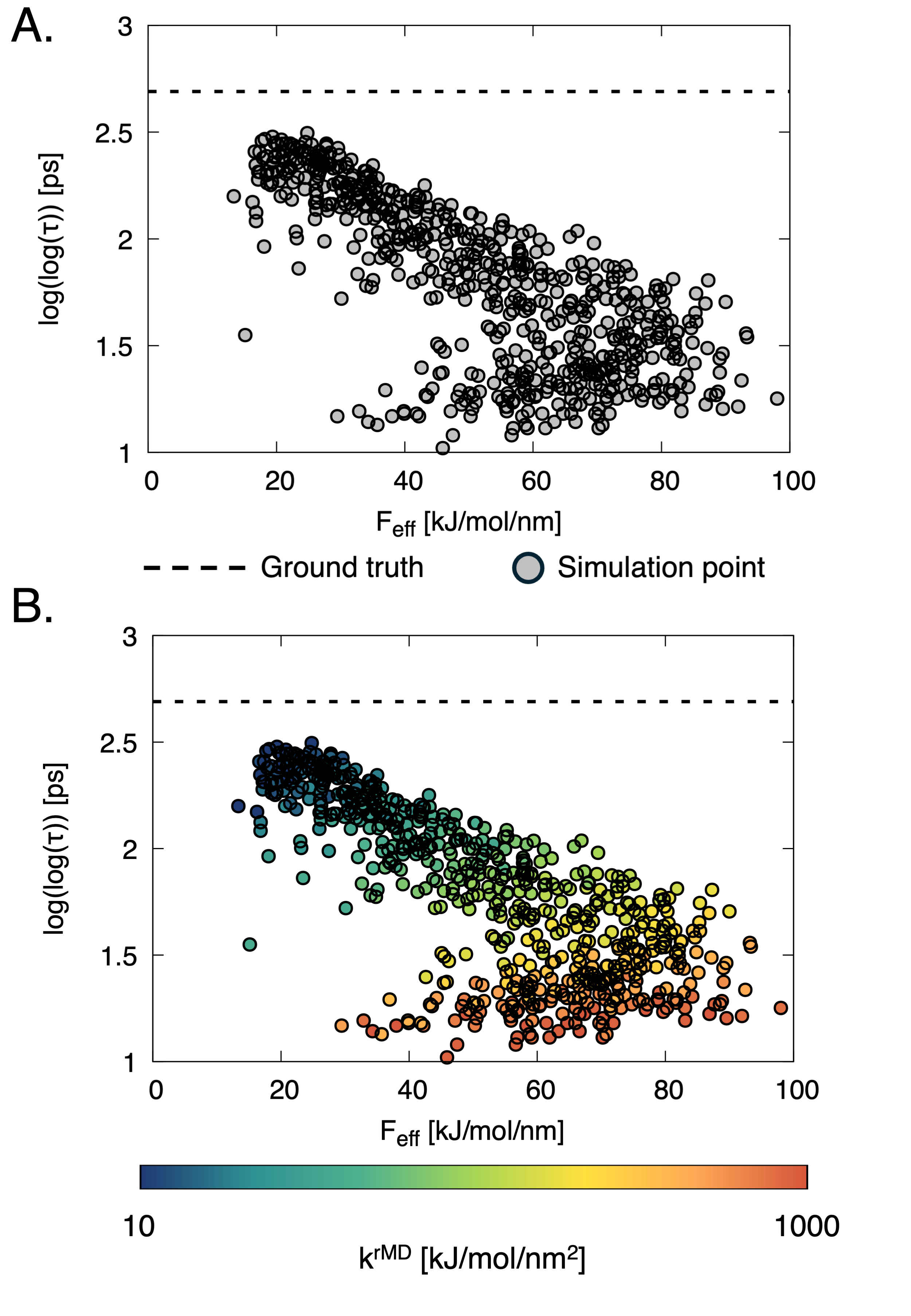}
    \caption{Distribution of the $\log(\log(\tau))$ as a function of $F_{\text{eff}}$ in multiple rMD runs at different $k^{\text{rMD}}$ with the multi-eGO model. In A) we can see the general distribution with the two regimes, adiabatic and inertial, for the transition. In B) we can observe how the transition times are distributed with respect to the applied $k^{\text{rMD}}$, showing that the majority of the outliers from the linear adiabatic distribution are coming from the high $k^{\text{rMD}}$ simulations.}
    \label{img:tau_mego}
\end{figure}

In our protocol, we performed 660 rMD simulations, with the harmonic constant $k^{\text{rMD}}$ ranging from 1 to 100 kJ/mol/nm$^2$ (20 replicas for each selected $k^{\text{rMD}}$ value -- the complete list is in Table S2 in the Supporting Information), for a total simulation time of $\sim 6.1\,\mu$s. The distribution of $\log(\log(\tau))$ as a function of $F_{\text{eff}}$ is shown in Figure \ref{img:tau_mego}.

Observing the distribution of $\log(\log(\tau))$ versus $F_{\text{eff}}$, we can see that there are two regimes: (i) a linear dependence, with larger $\tau$ having smaller $F_{\text{eff}}$, as expected from our framework, and (ii) a sparse distribution of points for small $\log(\log(\tau))$ and large $F_{\text{eff}}$. \\ We can interpret the two trends as the effect of two regimes of the transition: in the linear case, the bias drives the transition in an adiabatic way, while for the sparse one, there is an inertial effect where the ligand absorbs a large amount of energy (this happens for steep rMD potentials) and overcomes the barriers at an effective temperature much higher than the supposed simulation temperature $T$. 

\begin{figure}[h!]
    \includegraphics[width=0.45\textwidth]{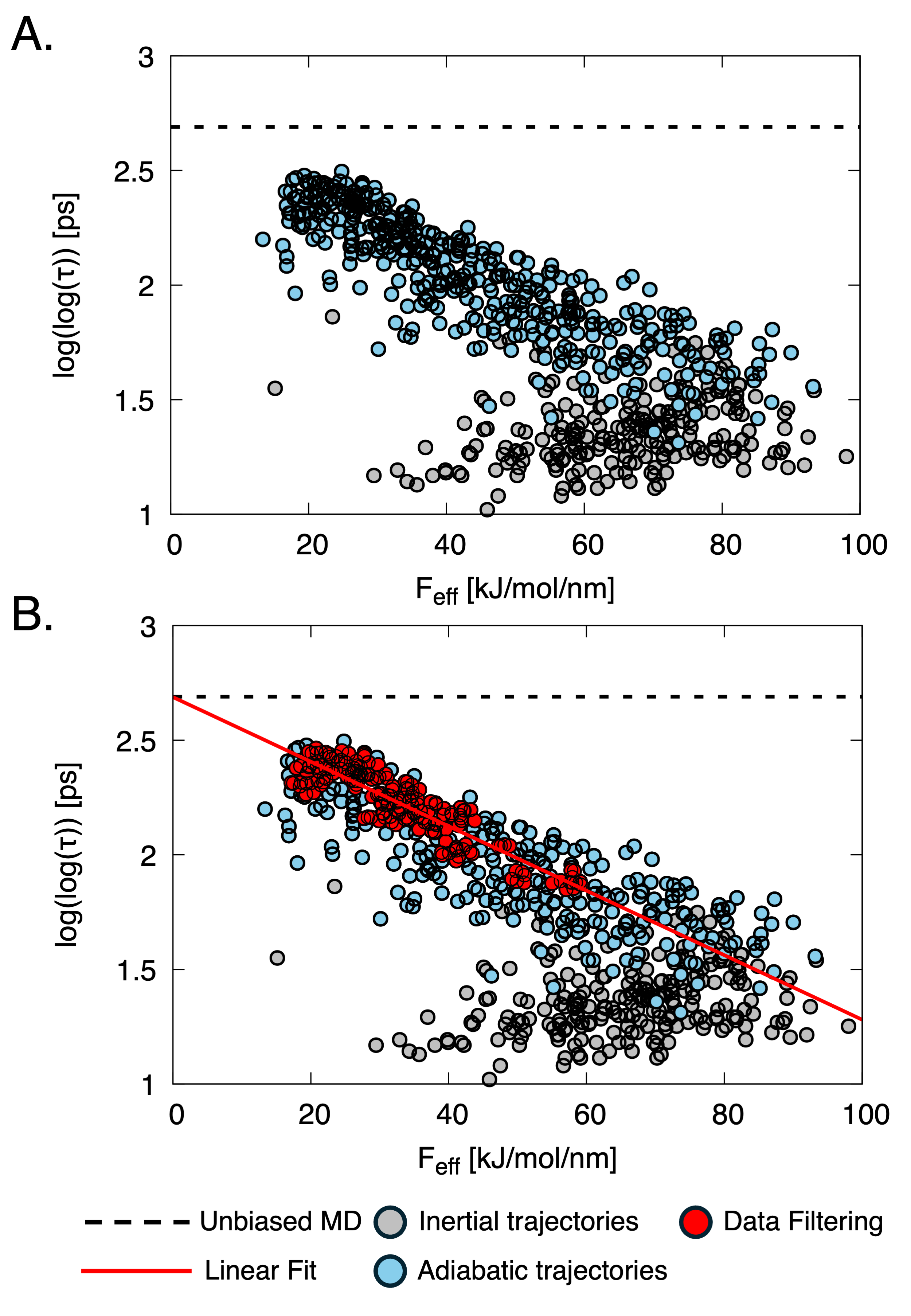}
    \caption{Results of clustering procedure and fitting for multi-eGO rMD runs. A) Separation between inertial (gray) and adiabatic (light blue) trajectories. B) Removal of sparse point and identification of relevant simulation points (red) with its fitted linear function (red). }
    \label{img:cluster_mego}
\end{figure}

To classify the trajectories between adiabatic and inertial ones, we performed an unsupervised clustering using the Density Peak Clustering\cite{Rodriguez2014} (DPC), having as input data the timeseries of the CV of the rMD trajectories (details in the Methods section). We imposed the presence of two clusters, obtaining a clear separation of the two observed trends in the $\log(\log(\tau))$/$F_{\text{eff}}$ plot (Figure \ref{img:cluster_mego}A).
However, the sparse nature of the points and the fact that we are fitting a line on a double exponential function makes the removal of outliers extremely important. The first DPC clustering cannot discriminate between inertial trajectories at small $k^{\text{rMD}}$ and adiabatic ones at large $k^{\text{rMD}}$ for the fast transition involved in both. We therefore decide to cluster again the remaining data using as input the 2D ($\log(\log(\tau))$, $F_{\text{eff}}$) data, imposing a single cluster and a variable data density threshold $\rho_{\text{thr}}$ to discriminate between cluster data and noise. To define the threshold (and use it also in cases where the final $k_{\text{off}}$ value is unknown), we adjusted its value to minimize the difference between the rMD-derived $k_{\text{off}}$ and the unbiased MD value, arriving at $\rho_{\text{thr}} = 0.3$ (details in Methods and Supporting Information). The final results are in Figure \ref{img:cluster_mego}B.

Numerically, we have $k^{\text{rMD}}_{\text{off}} = 4.1\cdot 10^{5}$ s$^{-1}$ ($[2.5\cdot 10^{5},6.6\cdot 10^{5}]$ s$^{-1}$ 95\% CI).

It is important to underline that the multi-eGO model here served solely to obtain the density threshold $\rho_{\mathrm{thr}}$ for the DPC filtering step, exploiting its ability to provide extensive unbinding statistics at low computational cost.

\subsection{All-atom Simulations}

We prepared our atomistic model starting from the crystallographic structure deposited in the Protein Data Bank (code: 3PTB\cite{Marquart1983}), performing a minimization and relaxation protocol to prepare the system (details in the Methods section).

\begin{figure*}[h!]
    \includegraphics[width=0.85\textwidth]{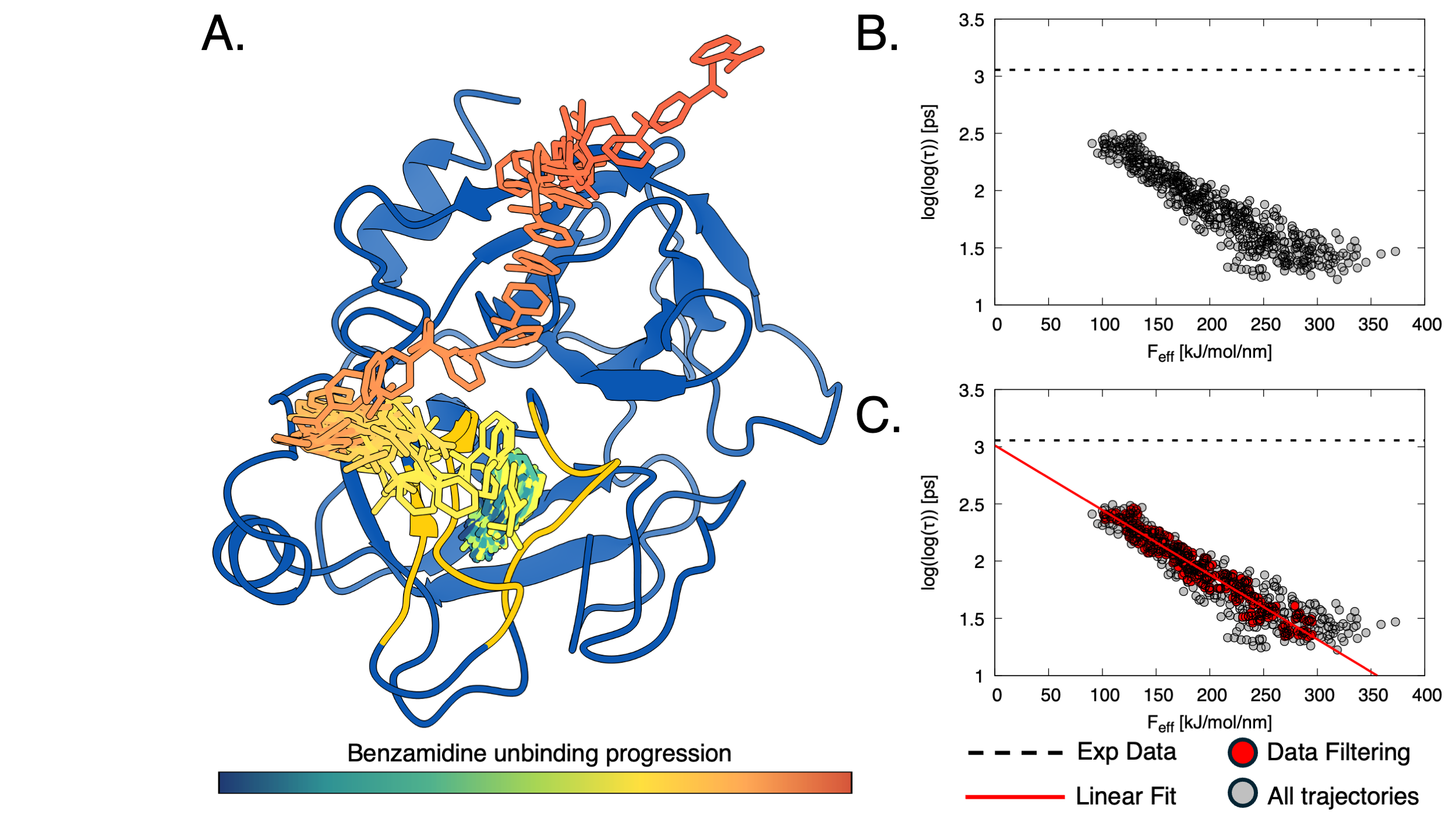}
    \caption{Results for the all-atom simulations. A) Example of an unbinding transition of benzamidine from trypsin; in blue the protein residues not in contact (farther than 5 \AA{} from the ligand in the bound structure), in yellow the pocket residues, and in rainbow palette the progression of benzamidine unbinding. B) Distribution of the $\log(\log(\tau))$ in function of $F_{\text{eff}}$ in multiple rMD runs at different $k^{\text{rMD}}$ with the all-atom model; we can observe the  absence of inertial trajectories. C) Removal of sparse point and identification of relevant simulation points (red) with its fitted linear function (red).}
    \label{img:panel_allatom}
\end{figure*}

In the rMD protocol we consider the same CV used for the multi-eGO model, i.e., the distance between the center of mass of benzamidine and the center of mass of the binding pocket residues C$_{\alpha}$ (the list of residues is in Table S1). For the values of $k^{\text{rMD}}$, we set the range from 16 to 1000 kJ/mol/nm$^{2}$, with 20 replicas per $k^{\text{rMD}}$ value, for a total number of 600 different runs (complete list of $k^{\text{rMD}}$ in Table S3). The total simulation time is $\sim 5.4 \mu$s. The results are in Figure \ref{img:panel_allatom}.

The first clear difference in the all-atom simulations (compared to multi-eGO) is the near complete absence of inertial trajectories. Virtually all runs follow a linear dependence between $\log(\log(\tau))$ and $F_{\text{eff}}$ (Figure \ref{img:panel_allatom}B). We attribute this to the presence of explicit solvent, which absorbs most of the excess kinetic energy given by the rMD potential. For this reason, thus we skipped the first, inertial trajectory filter and proceeded directly to remove sparse outliers before fitting (Figure \ref{img:panel_allatom}C).

From the linear fit and the bootstrapping procedure, we get a $k^{\text{rMD}}_{\text{off}} = 1.5\cdot 10^{3}$ s$^{-1}$ ($[0.8\cdot 10^{3},2.6\cdot 10^{3}]$ s$^{-1}$ 95\% CI), in agreement with the experimental value $k^{\text{exp}}_{\text{off}}=(6 \pm 3) \cdot 10^{2}\,\text{s}^{-1}$. 

\begin{figure}[h!]
    \includegraphics[width=0.45\textwidth]{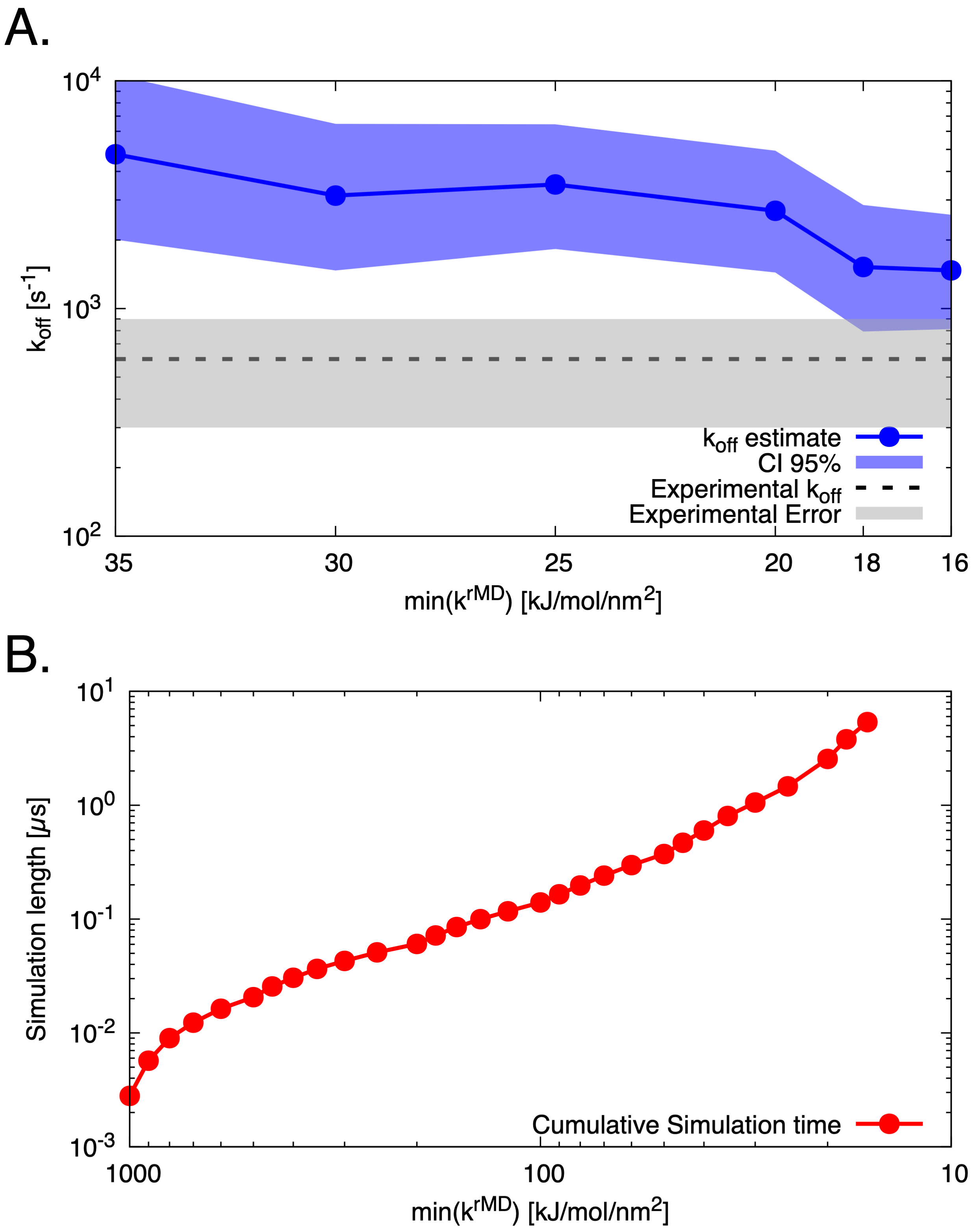}
    \caption{Efficacy and efficiency analysis of the approach. A) obtained $k_{\text{off}}$ values as a function of the minimum value of $k^{\text{rMD}}$ compared with the experimental value. B) Cumulative simulation time as a function of the minimum value of $k^{\text{rMD}}$; our approach scales (at least) exponentially with the $k^{\text{rMD}}$ value.}
    \label{img:efficiency}
\end{figure}

Finally, we estimated the computational cost and convergence properties of the approach to assess its efficiency and applicability.
In particular, we analyze how the final $k_{\text{off}}$ value changes as a function of the minimum value of $k^{\text{rMD}}$ (Figure \ref{img:efficiency}A) and the amount of simulation time needed to compute all the simulations from the maximum value of $k^{\text{rMD}}$ to a lower bound.

We can see that the order of magnitude of the $k_{\text{off}}$ value is matched with a $\min(k^{\text{rMD}})=30$ kJ/mol/nm$^{2}$, which corresponds to a total simulation time of around $1\,\mu$s. The matching of the experimental and computational distribution (i.e., when there is an overlap between 95\% CI of rMD value and the experimental error) is reached with $\min(k^{\text{rMD}})=18$ kJ/mol/nm$^{2}$, for a total simulation time of $\sim3.8\,\mu$s, which is in line with the state-of-the-art in the field of kinetic rates calculation for protein-ligand binding.

\section{\label{conclusion}Conclusions}
We presented a non-equilibrium approach based on rMD simulation to quantitatively estimate kinetic rates, applying it to a protein-ligand system. Refining the technique on implicit solvent model and applying it on an explicit solvent models for benzamidine-trypsin demonstrates that our approach can accurately reproduces experimental kinetics with competitive computational costs. In our case, identifying a CV capable of effectively describing the bound, unbound, and intermediate states was straightforward, as it had already been validated in previous studies on the same system. More generally, however, the selection of an appropriate CV is a critical step for this class of enhanced sampling methods, as it can strongly influence both efficiency and accuracy, and should therefore be carried out with particular care. We underline the fact that, despite being used here for a ligand unbinding problem, this approach can be used in principle to address any kind of kinetic rate calculation, given the knowledge of an apt CV. 
Regarding the applicability of the approach, the value of $\rho_{\mathrm{thr}}$ was determined using the simplified multi-eGO model, and we consider it transferable to any non-equilibrium process with a comparable energy scale. This follows from its dependence solely on the statistical distribution of transition times and forces, eliminating the need for recalibration.
The adaptability of the method suggests potential application in various molecular systems, positioning rMD as a versatile and powerful tool for future kinetic rate investigations beyond protein-ligand systems.


%
%

%

\begin{acknowledgments}
We thank Carlo Camilloni and Guido Tiana for useful discussions, Michele Invernizzi and Paolo Carloni for the critical reading of the manuscript.
This research was supported by Universit\`a degli Studi di Milano (Piano di supporto di ateneo Linea 2 2023-DBS Capelli).
\end{acknowledgments}

\section*{Data Availability Statement}
All the input data, trajectories, CV progression, and analysis scripts are openly available in
Zenodo at http://doi.org/10.5281/zenodo.15224958, reference number 15224958. \\
The input files are deposited in the PLUMED-NEST\cite{plumed2019} public repository, ID 25.010.

\bibliography{biblio}

\end{document}



\title{Supporting information for: ``Kinetic rates calculatio
n via non-equilibrium dynamics''} 

\author{Bruno Stegani}%
\affiliation{ 
Department of Biosciences, Università degli Studi di Milano, Via Celoria 26, 20133 Milan, Italy.
}

\author{Riccardo Capelli}%
 \email{riccardo.capelli@unimi.it}
\affiliation{ 
Department of Biosciences, Università degli Studi di Milano, Via Celoria 26, 20133 Milan, Italy.
}

\date{\today}


\pacs{}

\maketitle 

\section{Multi-eGO force field preparation}
To construct the multi-eGO apo potential, we reweighted our prior simulation --consisting of the apo protein with only bonded interactions-- using the contact probabilities obtained from the training simulation.

\begin{figure}
    \includegraphics[width=0.6\textwidth]{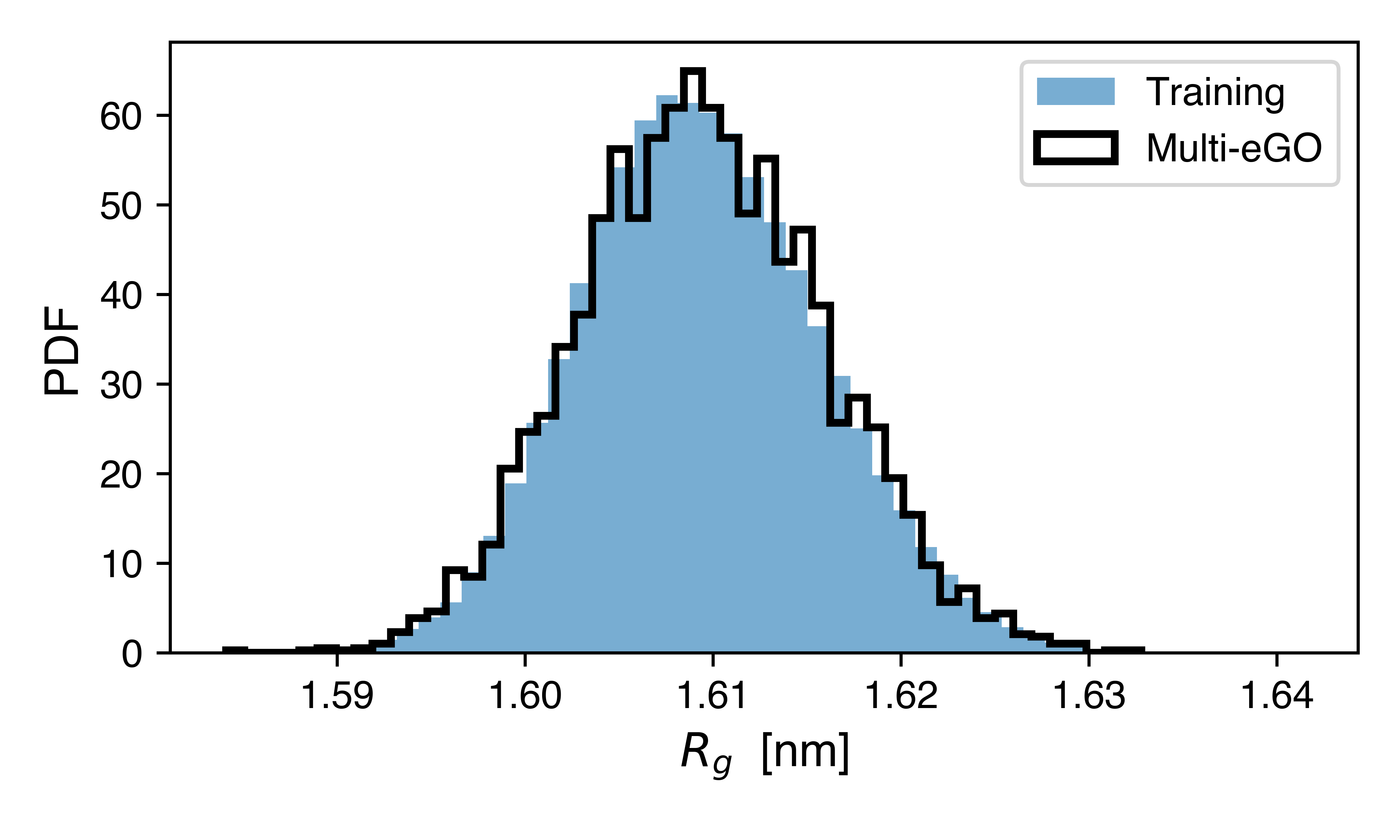}
    \caption{Comparison of the radius of gyration ($R_g$) distribution between the training simulation (blue) and the multi-eGO simulation (black), using the optimal intra-molecular parameter $\varepsilon_0^{\text{intra}} = 0.35$ kJ/mol. }
    \label{SIimg:gyration_apo}
\end{figure}

\begin{figure}
    \includegraphics[width=\textwidth]{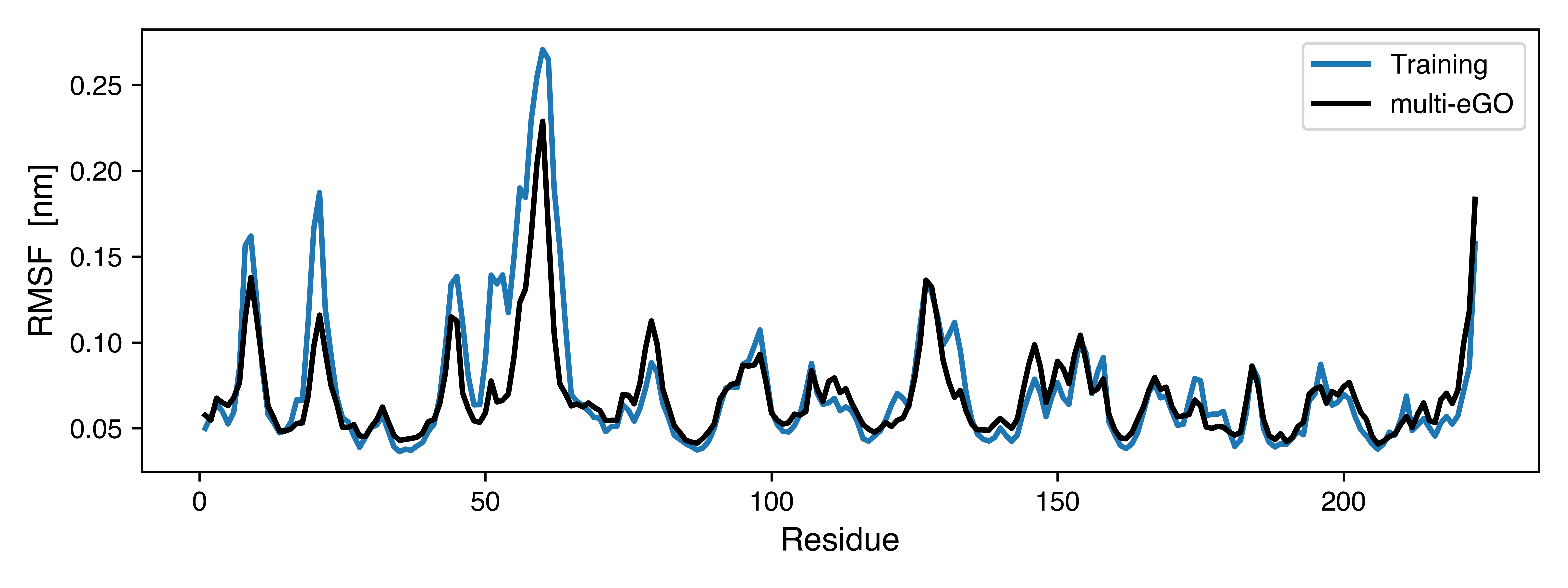}
    \caption{Comparison of residue-wise RMSF between the training simulation (blue) and the multi-eGO simulation (black), using the optimal intra-molecular parameter $\varepsilon_0^{\text{intra}} = 0.35$ kJ/mol. }
    \label{SIimg:rmsf_apo}
\end{figure}

To define the free parameter $\varepsilon_0^{\text{intra}}$, we performed a series of simulations at different values of $\varepsilon_0^{\text{intra}}$, comparing key structural properties between the multi-eGO and training simulations. Specifically, we considered the radius of gyration ($R_g$) and the root mean square fluctuation (RMSF), selecting the $\varepsilon_0^{\text{intra}}$ value that yielded the best agreement. Figures \ref{SIimg:gyration_apo} and \ref{SIimg:rmsf_apo} show the comparison of the $R_g$ distribution and residue-wise RMSF, respectively, for the optimal value of $\varepsilon_0^{\text{intra}} = 0.35$ kJ/mol.

The holo multi-eGO potential was similarly obtained by reweighting the training contact probabilities using those from a corresponding inter-molecular prior. This prior consisted of a simulation at the same ligand concentration, where solvated ligands interacted with the folded protein solely via excluded volume. The resulting contact probabilities capture random ligand-protein clashes at the correct concentration, enabling reweighting that removes the contribution of rototranslational entropy.

\begin{figure}
    \includegraphics[width=0.6\textwidth]{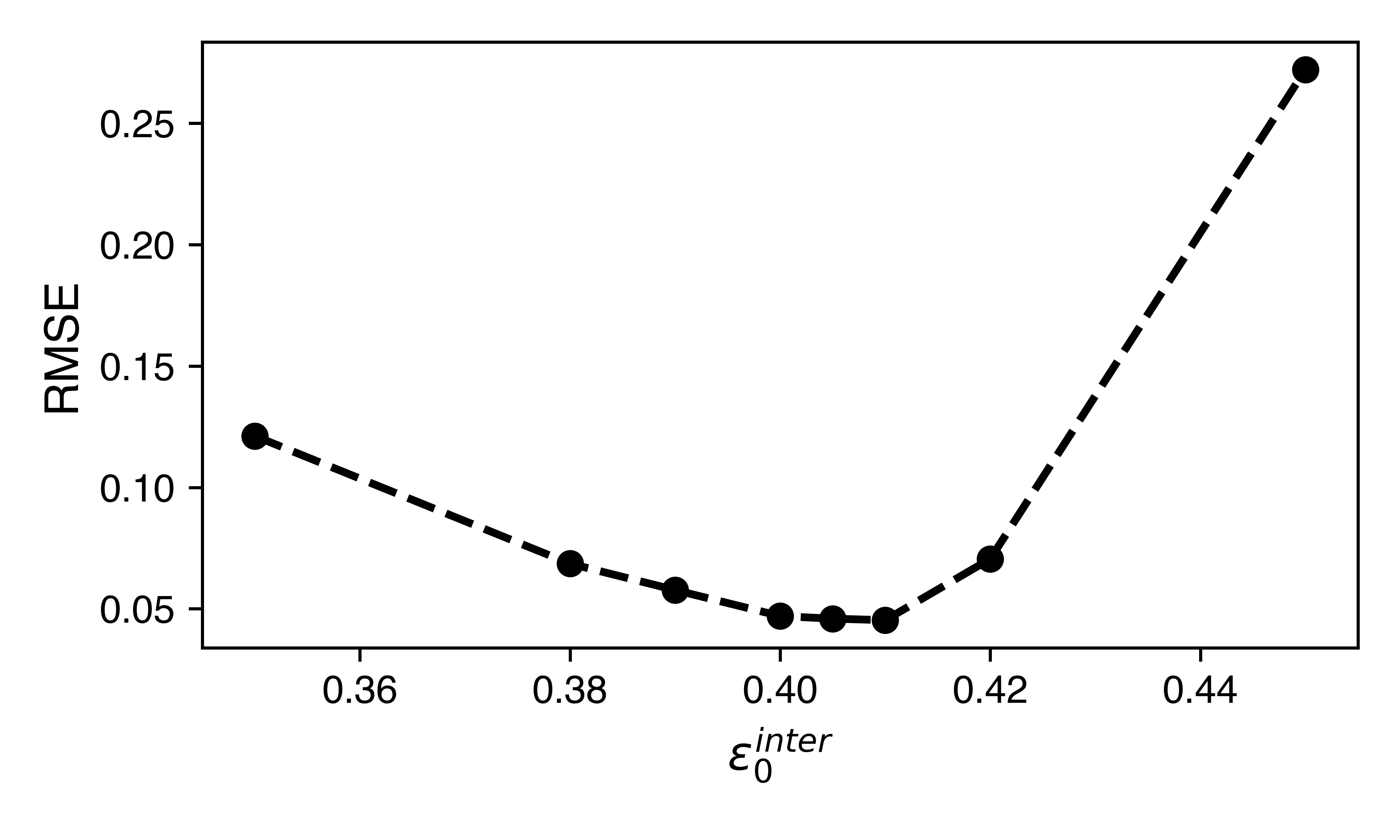}
    \caption{Root mean square error (RMSE) between the training and multi-eGO inter-molecular residue-wise contact maps as a function of the inter-molecular parameter $\varepsilon_0^{\text{inter}}$. The best agreement is observed at $\varepsilon_0^{\text{inter}} = 0.41$ kJ/mol, corresponding to the minimum RMSE.}
    \label{SIimg:rmse}
\end{figure}

To determine the optimal value of $\varepsilon_0^{\text{inter}}$, we compared inter-molecular residue contact maps across values ranging from 0.35 to 0.45 kJ/mol. We quantified the agreement using the root mean square error (RMSE), defined as $RMSE = \sqrt{\frac{1}{N}\sum_i (p_i^{\text{train}}-p_i^{\text{mego}})^2}$ where $p_i$ denotes the contact probability of residue pair $i$ in the training and multi-eGO simulations. Figure \ref{SIimg:rmse} shows the RMSE as a function of $\varepsilon_0^{\text{inter}}$, with the minimum at 0.41 kJ/mol, which we selected as the optimal value.
\begin{figure}
    \includegraphics[width=0.8\textwidth]{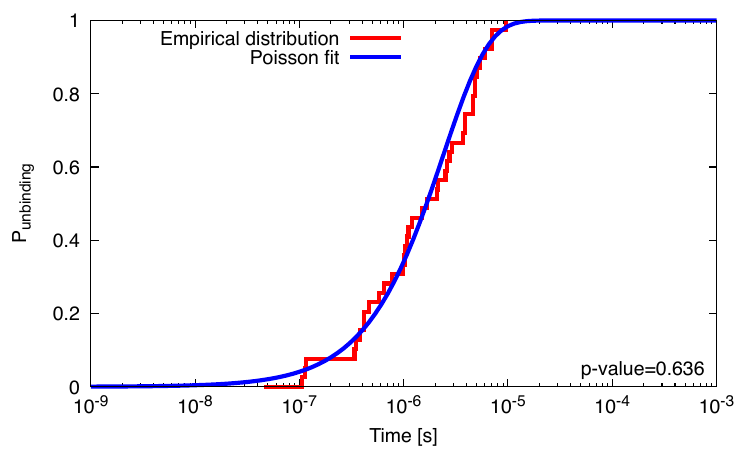}
    \caption{Comparison of the empirical distribution of residence times observed via unbiased multi-eGO MD (red) and a fitted Poisson distribution (blue). Kolmogorov-Smirnov test between the two distributions resulted in a p-value of 0.636, which suggest (as expected) that the unbinding process follows a rare event distribution.}
    \label{SIimg:poisson_unbiased}
\end{figure}

\begin{figure}
    \includegraphics[width=0.8\textwidth]{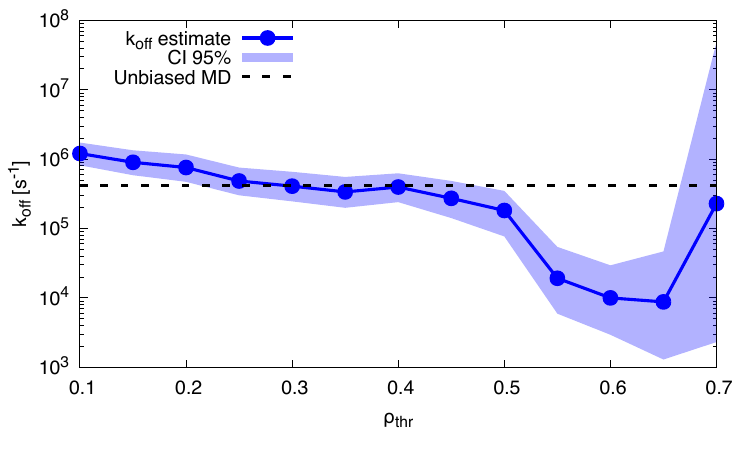}
    \caption{Change in $k_{\text{off}}$ value in function of the choice of the clustering parameter $\rho_{\text{thr}}$ in multi-eGO simulations. Given the monotonic decrease of the value of $k_{\text{off}}$ with $\rho_{\text{thr}}$, we choose to set the value of $\rho_{\text{thr}}=0.3$, which is the first crossing of the expected value. }
    \label{SIimg:thr_choice}
\end{figure}

\begin{figure}
    \includegraphics[width=\textwidth]{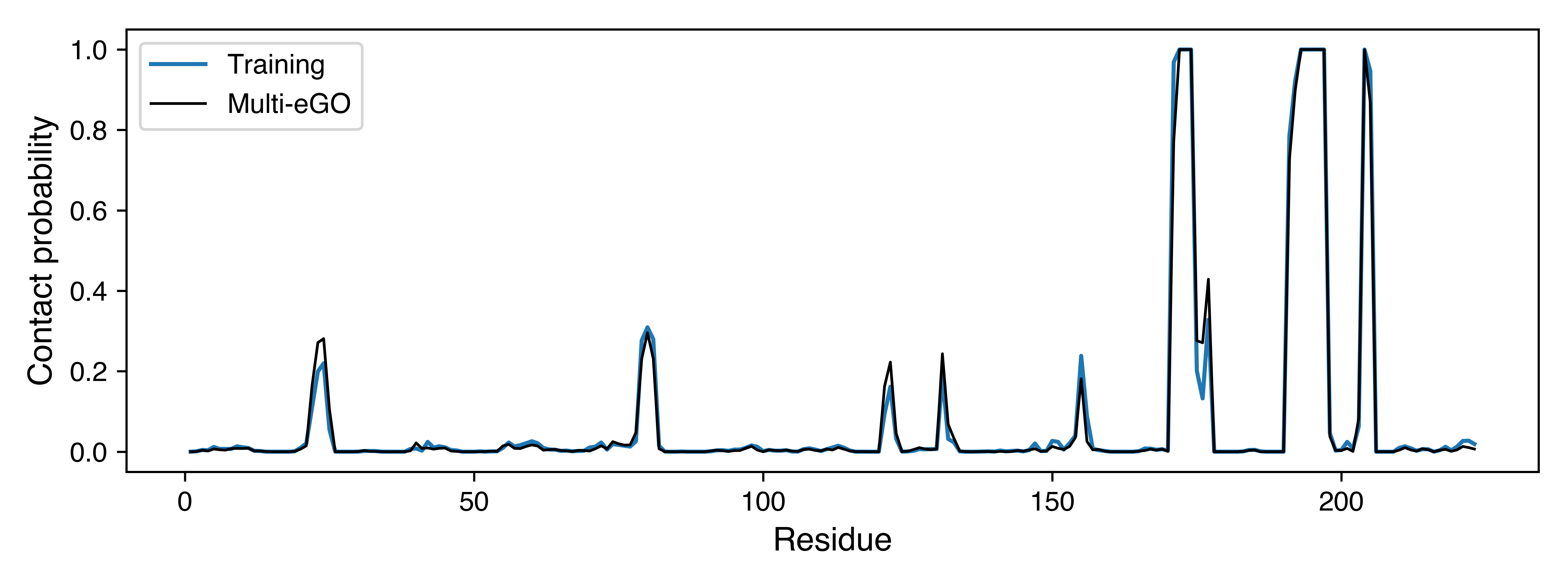}
    \caption{Comparison of the inter-molecular residue-wise contact map between Trypsin and Benzamidine from the training simulation (blue) and the multi-eGO simulation (black), using the optimal parameter $\varepsilon_0^{\text{inter}} = 0.41$ kJ/mol.  }
    \label{SIimg:contact_holo}
\end{figure}
%

\clearpage

\begin{table}
\caption{\label{tab:pocket_res} List of the residues which define the binding pocket.}
\begin{tabular}{c |c}
\multicolumn{2}{c}{Residues} \\
\hline
ASP171  & GLY194 \\
SER172  & SER195 \\
CYS173  & GLY196 \\
GLN174  & CYS197 \\
SER177  & ALA198 \\
VAL191  & PRO203 \\
SER192  & GLY204 \\
TRP193  & VAL205 \\
\end{tabular}
\end{table}

\begin{table}
\caption{\label{tab:mego_sim} rMD simulations performed using the multi-eGO potential.}
\begin{tabular}{c|c||c|c}
Replica IDs & $k$ $\left[\frac{kJ}{\text{mol}\cdot\text{nm}^2}\right]$] & Replica IDs & $k$ $\left[\frac{kJ}{\text{mol}\cdot\text{nm}^2}\right]$\\
\hline
1-20    & 100   & 341-360 & 9   \\
21-40   & 90    & 361-380 & 8   \\
41-60   & 80    & 381-400 & 7   \\
61-80   & 70    & 401-420 & 6   \\
81-100  & 60    & 421-440 & 5   \\
101-120 & 50    & 441-460 & 4.5 \\
121-140 & 45    & 461-480 & 4   \\
141-160 & 40    & 481-500 & 3.5 \\
161-180 & 35    & 501-520 & 3   \\
181-200 & 30    & 521-540 & 2.5 \\
201-220 & 25    & 541-560 & 2   \\
221-240 & 20    & 561-580 & 1.8 \\
241-260 & 18    & 581-600 & 1.6 \\
261-280 & 16    & 601-620 & 1.4 \\
281-300 & 14    & 621-640 & 1.2 \\
301-320 & 12    & 641-660 & 1 \\
321-340 & 10    &    &  \\
\end{tabular}
\end{table}

\begin{table}
\caption{\label{tab:mego_sim} rMD simulations performed using the all-atom potential.}
\begin{tabular}{c|c||c|c}
Replica IDs & $k$ $\left[\frac{kJ}{\text{mol}\cdot\text{nm}^2}\right]$] & Replica IDs & $k$ $\left[\frac{kJ}{\text{mol}\cdot\text{nm}^2}\right]$\\
\hline
1-20    & 1000  & 301-320 & 120   \\
21-40   & 900   & 321-340 & 100   \\
41-60   & 800   & 341-360 & 90   \\
61-80   & 700   & 361-380 & 80   \\
81-100  & 600   & 381-400 & 70   \\
101-120 & 500   & 401-420 & 60  \\
121-140 & 450   & 421-440 & 50  \\
141-160 & 400   & 441-460 & 45  \\
161-180 & 350   & 461-480 & 40  \\
181-200 & 300   & 481-500 & 35  \\
201-220 & 250   & 501-520 & 30  \\
221-240 & 200   & 521-540 & 25 \\
241-260 & 180   & 541-560 & 20 \\
261-280 & 160   & 561-580 & 18 \\
281-300 & 140   & 581-600 & 16 \\
\end{tabular}
\end{table}

\section{Theory}
We add here the mathematical steps to obtain the double logarithmic dependency of $\tau$ as a function of $\tau_0$.
Starting from 
\begin{equation}
    \tau = \tau_0 \exp \left( -\frac{\gamma_0 F_c}{k_{B}T} \left( 1- \exp \left( -\frac{F_{\text{eff}}}{F_{c}} \right) \right) \right)
\end{equation}
and taking the first logarithm yelds:
\begin{equation}\label{eq:1log}
    \log(\tau) = \log(\tau_0) -\frac{\gamma_0 F_c}{k_{B}T} \left( 1- \exp \left( -\frac{F_{\text{eff}}}{F_{c}} \right) \right)
\end{equation}
taking now the second logarithm, we obtain
\begin{equation}\label{eq:2log_a}
    \log(\log(\tau)) = \log\left(\log(\tau_0)-\frac{\gamma_0 F_c}{k_{B}T} \left( 1- \exp \left( -\frac{F_{\text{eff}}}{F_{c}} \right)\right)\right)
\end{equation}
which can be rewritten as
\begin{equation}\label{eq:2log_b}
    \log(\log(\tau)) = \log\left(\log(\tau_0)\left( 1 - \frac{\gamma_0 F_c}{k_{B}T \log(\tau_0)} \left( 1- \exp \left( -\frac{F_{\text{eff}}}{F_{c}} \right)\right)\right)\right)
\end{equation}
yielding the same double-logarithmic dependence as shown in the main text
\begin{equation}
    \log(\log(\tau)) = \log(\log(\tau_0))+ \log\left( 1 - \frac{\gamma_0 F_c}{k_{B}T \log(\tau_0)} \left( 1- \exp \left( -\frac{F_{\text{eff}}}{F_{c}} \right)\right)\right)
\end{equation}
